\documentclass[11pt]{article}

\usepackage{amsmath,amssymb,graphicx}

\def\mydate{October 19, 2006}

\newcounter{sxn}

\newcounter{axn}

\date{}

\def\ignore#1{{}}

\newcommand{\beeq}{\begin{equation}}
\newcommand{\eneq}{\end{equation}}
\newcommand{\beqn}{\begin{eqnarray}}
\newcommand{\eeqn}{\end{eqnarray}}

\def\dd{\partial}
\def\la{\raise.16ex\hbox{$\langle$}\lower.16ex\hbox{}  }
\def\ra{\, \raise.16ex\hbox{$\rangle$}\lower.16ex\hbox{} }
\def\go{\rightarrow}

\def\onehalf{ \hbox{${1\over 2}$} }
\def\half{ \frac{1}{2} }

\def\eff{{\rm eff}}

\def\EM{{\rm EM}}
\def\BC{{\rm BC}}
\def\phys{{\rm phys}}
\def\diag{{\rm diag}~ }

\def\ep{\epsilon}
\def\psibar{ \psi \kern-.79em\raise.6em\hbox{$-$}\, }
\def\psibarl{ \psi \kern-.65em\raise.6em\hbox{$-$} \lower.6em\hbox{} }

\def\mypmatrix#1{\begin{pmatrix} #1 \end{pmatrix}}
\def\mybmatrix#1{\begin{bmatrix} #1 \end{bmatrix}}

\begin{document}

\thispagestyle{empty}

\baselineskip=11pt

{\small    \hfill OU-HET 566/2006}

{\small    \hfill \mydate}

\vskip .7cm

\begin{center}
{\LARGE \bf Dynamical Gauge-Higgs Unification\footnote{Proceeding for the
Cairo International Conference on High Energy Physics, 
German University in Cairo, Egypt, 14 - 17 January 2006.}}

\vskip .5cm 
{\large  Yutaka Hosotani}\\
 \vskip .5cm

{\small \it Department of Physics, Osaka University, Toyonaka, 
Osaka 560-0043, Japan}\\
\end{center}

\begin{abstract}
{\small
In the dynamical gauge-Higgs unification  the 4D Higgs field is unified with 
gauge fields and the electroweak symmetry is dynamically broken 
by the Hosotani mechanism.  Interesting phenomenology 
is obtained in the Randall-Sundrum warped spacetime.
(i) The Higgs boson mass is predicted at the LHC energies.
(ii) The hierarchy in the fermion mass spectrum is naturally explained.
(iii) Tiny violation of the universality in the charged current interactions is
predicted.  (iv) Yukawa couplings of quarks and leptons are suppressed 
compared with those in the standard model.  (v) $WWH$, $ZZH$ 
couplings  are suppressed compared with those in the standard model.
}
\end{abstract}


\vskip .5cm

\baselineskip=15pt

\section{Introduction}

In the gauge-Higgs unification   4D Higgs scalar fields are unified with 4D gauge 
fields within the framework of higher dimensional gauge theory.   
Low energy modes of  extra-dimensional components of gauge potentials 
are 4D Higgs fields.  The scenario works remarkably well when the 
extra-dimensional space is non-simply-connected.\cite{YH1, YH2}  There arise Yang-Mills
AB (Aharonov-Bohm) phases along the extra dimension,  whose fluctuations
in four dimensions are nothing but the 4D Higgs fields.  The most notable feature 
is that quantum dynamics generate non-trivial, finite effective potential for the 
Higgs fields, inducing dynamical gauge symmetry breaking  and generating
finite masses for the Higgs fields at the same time.  Even though the theory is
non-renormalizable,  many properties of the Higgs fields can be deduced 
irrespective of unknown dynamics at the cutoff scale.  The Higgs boson mass 
turns out to be much smaller than the Kaluza-Klein mass scale.  This is contrasted
to the earlier proposal of the gauge-Higgs unification based on the ad hoc
symmetry ansatz.\cite{Fairlie1, Manton1}

In the last ten years the scenario of the gauge-Higgs unification  has been 
applied to the electroweak interactions and grand unified theories with the aid of
orbifolds as extra dimensions.\cite{Lim2}-\cite{SH} 
In this article we focus on applications to the electroweak interactions 
where, besides the Higgs boson mass, many illuminating predictions 
are made for LHC and linear colliders.  

To achieve the gauge-Higgs unification in the electroweak interactions,
there are a few requirements to be fulfilled.  First of all,  the electroweak gauge
symmetry is $SU(2)_L \times U(1)_Y$ which breaks down to $U(1)_\EM$
triggered by non-vanishing VEV of an $SU(2)_L$ doublet Higgs field. 
 In order for the 4D Higgs field
be a part of gauge potentials the original gauge group must be larger than 
$SU(2)_L \times U(1)$.    Secondly, fermion content must be chiral.  The 
second requirement is restrictive, as fermions in higher dimensions
tend to lead to vectorlike theory in the effective 4D theory at low energies,
unless the extra-dimensional space has nontrivial topology or 
there exists nonvanishing flux in the extra dimensions.  These requirements
can be naturally and easily fulfilled when the extra dimensional space is
an orbifold.

\subsection{Gauge theory on an orbifold}

Consider $S^1$ with a coordinate $y$ where $y$ and $y+2\pi R$ are identified.
Further we identify $y$ and $-y$, which gives an orbifold $S^1/Z_2$.  
There appear two fixed points under parity; $y_0=0$ and $y_1=\pi R$.  
Let us analyse gauge theory on $M^4 \times (S^1/Z_2)$, which is first defined
on a covering space $M^5$, supplemented with restrictions appropriate to
preserve the nature of $S^1/Z_2$.  Although $y_j + y$ and $y_j -y$ represent the same 
physical point, gauge potentials need not be the same.  They may differ from
each other up to a gauge transformation.  The orbifold structure is respected if
\beeq
\begin{pmatrix} A_\mu \cr A_y  \end{pmatrix} (x, y_j -y) =
P_j \mypmatrix{A_\mu \cr - A_y}  (x, y_j + y) P_j^\dagger   
\label{BC1}
\eneq
where $P_j$ is an element of the gauge group satisfying $P_j^2 = I$.  Similarly 
for fermions in the spinor representation in an $SU(N)$ group or in the 
vector representation in an $SO(N)$ group
\beeq
\psi(x, y_j - y) = \pm P_j \gamma^5 \psi(x, y_j + y) ~~.
\label{BC2}
\eneq

If $P_j \not{\kern -2pt \propto} ~ I$, the gauge symmetry $G$  apparently 
breaks down to a
smaller subugroup $H_{\BC}$.   $\{ P_0, P_1 \}$ defines the symmetry of boundary condition.
$H_{\BC}$ is not necessarily the physical symmetry $H_\phys$ which survives at the end.  
$H_\phys$ can be either smaller or larger than $H_{\BC}$.  Put it differently,
two distinct sets of boundary conditions, $\{ P_0, P_1 \}$ and $\{ P_0', P_1' \}$
can be equivalent to each other in physics content.  All of these are due to dynamics of
Yang-Mills AB phases.    It is called the Hosotani mechanism.\cite{YH1, YH2, HHHK}
In the application to
the electroweak interactions we would like to have $H_{\BC} = SU(2)_L \times  U(1)_Y$
and $H_\phys = U(1)_\EM$.

In the $SU(3)$ model,  $P_0=P_1=\diag (-1, -1,1)$ gives $H_{\BC} = SU(2) \times  U(1)$.
Zero modes exist for the $H_{\BC}$ part of $A_\mu$ and for the $G/H_{\BC}$ part of $A_y$
which forms an $SU(2)$ doublet and idetified with the 4D Higgs field.   Although this 
model gives an incorrect Weinberg angle, it gives a nice working ground to investigate
physics of the $W$ boson and fermions.  Another model of interest is the 
$SO(5) \times U(1)_{B-L}$ model proposed by Agashe et al.\cite{Agashe2}   
For the $SO(5)$ part we
take $P_0=P_1= \diag (-1,-1,-1,-1,1)$, which gives 
$H_{\BC}' = SO(4) \times U(1)_{B-L} = SU(2)_L \times SU(2)_R \times U(1)_{B-L}$.
With additional dynamics on the one of the branes at $y=0$, the symmetry of 
boundary conditions is reduced to $H_{\BC} = SU(2)_L \times  U(1)_Y$.
Zero modes of $A_y$ are located at
\beeq
A_y \sim 
\mypmatrix{ &&&& -i \phi_1\cr &&&& -i \phi_2\cr &&&& -i \phi_3\cr &&&& -i \phi_4\cr
    i  \phi_1 &  i \phi_2 & i  \phi_3 & i  \phi_4 &}
      ~~,~~
\Phi = \mypmatrix{\phi_1 + i \phi_2 \cr \phi_4 - i \phi_3} ~~.
\label{Higgs1}
\eneq
$\Phi$ is the 4D Higgs doublet in the standard model.  We note that with the given
$\{ P_0, P_1 \}$,  $A_M$ becomes periodic; $A_M(x,y+2\pi R)= A_M(x,y)$.

\subsection{Yang-Mills AB phase $\theta_H$}

The zero modes of $A_y$ lead to non-Abelian generalization of the Aharonov-Bohm
phases (Yang-Mills AB phases).\footnote{In the literature they are often called  
Wilson line phases.}  The configuration gives vanishing field strengths $F_{MN}=0$,
but gives nontrivial phases
\beeq
e^{i \Theta_H/2} = P \exp \bigg\{ ig_A \int_0^{\pi R} dy \, A_y \bigg\}~~.
\label{ABphase1}
\eneq
The spectrum of gauge fields and fermions depends on  $\Theta_H$.
The phase $\Theta_H$ is a physical quantity.  As seen in eq.\  (\ref{Higgs1}),
the 4D Higgs fields are four-dimensional fluctuations of the Yang-Mills AB
phases.  This property leads to the finiteness of the Higgs boson mass.\cite{YH1}, 
\cite{Lim2}, \cite{YHscgt2}-\cite{YHfinite}

In the $SO(5) \times U(1)_{B-L}$ model one can suppose with the use of the residual
$SU(2)_L$ symmetry that only  the $ \phi_4$  
component of $A_y$ is nonvanishing in the vacuum.   The Yang-Mills AB 
phase $\theta_H$ is given by
\beeq
\Theta_H = \theta_H \cdot \Lambda ~~~,~~~ 
\Lambda = \mypmatrix{ 0\cr &0 \cr && 0 \cr &&&&-i \cr &&&i }
\eneq
There exist large gauge transformations which shift $\theta_H$ by multiples of $2\pi$,
while preserving the boundary conditions;
\beqn
&&\hskip -1cm
A_M' = \Omega A_M \Omega^\dagger + \frac{i}{g} \Omega \dd_M \Omega^\dagger 
~~,~~
\Omega = e^{iny/R \cdot \Lambda} \cr
\noalign{\kern 5pt}
&&\hskip -1cm
\theta_H ' = \theta_H + 2\pi n ~~.
\label{largeGT1}
\eeqn
It is seen that the phase nature of $\theta_H$ is a consequence of the large gauge 
invariance.

\section{Difficulties in flat space}

Before going into  detailed discussions in the Randall-Sundrum warped spacetime, 
we briefly summarize difficulties one encounters in gauge-Higgs unification
in flat spacetime.  The value of $\theta_H$ is dynamically determined  once the 
matter content is specified.  In typical situation the global minimum of the 
effective potential $V_\eff(\theta_H)$ is located either at $\theta_H=0$ or
at $\theta_H=O(1)$.  In the former case the gauge symmetry $H_\BC$ is unbroken, 
whereas in the latter case the  symmetry  breaks down to $U(1)_\EM$.  

The $W$ boson mass, $m_W$, becomes non-vanishing for $\theta_H \not= 0$ at
the tree level.   In flat space 
\beeq
m_W \sim  \frac{|\sin \theta_H |}{\pi R}
\label{Wmass1}
\eneq
which implies that the Kaluza-Klein mass scale $M_{KK}$ is too low.  Since the 
4D Higgs boson corresponds to four-dimensional fluctuations of $\theta_H$, its
mass arises as radiative corrections.  It turns out finite, but is given by
\beeq
m_H \sim \sqrt{ \frac{\alpha_W}{30} } ~ \frac{1}{R}  
\sim  \sqrt{ \frac{\alpha_W}{30} } ~ \frac{\pi m_W}{|\sin \theta_H |}
\label{Hmass1}
\eneq
which typically gives too small $m_H \sim 10 \, $GeV.  Of course, $\theta_H$ can
be small as a result of cancellations among contributions from various matter fields.  
However, it requires tuning of the matter content.  We argue that natural
resolution of the problem can be found once the gauge-Higgs unification is
achieved in the Randall-Sundrum spacetime.

\section{The Randall-Sundrum warped spacetime}

The Randall-Sundrum (RS) warped spacetime is given by 
\beqn
&&\hskip -1cm
ds^2 = e^{2\sigma(y)} ( \eta_{\mu\nu} dx^\mu dx^\nu - dy^2) \cr
\noalign{\kern 5pt}
&&\hskip -1cm
\sigma(y+2\pi R) = \sigma(y) = \sigma(-y) ~~, \cr
\noalign{\kern 5pt}
&&\hskip -1cm
\sigma(y) = k|y| \quad {\rm for~} |y| \le \pi R ~~,
\eeqn
where $\eta_{\mu\nu} = \diag (1,-1,-1,-1)$.  It has topology of $S^1/Z_2$.  
In the bulk five-dimensional spacetime $0<y<\pi R$, the cosmological 
constant is given by $- k^2$.  In other words, the RS spacetime is the
5D anti-de Sitter space sandwiched by the Planck brane (at $y=0$) and
the TeV brane (at $y= \pi R$).  The warp factor $e^{\pi kR}$ provides 
natural explanation of the large hierarchy factor
$M_{\rm Pl}/m_W \sim 10^{17}$, as was originally pointed out by Randall
and Sundrum.\cite{RS1}
We examine gauge theory defined on the RS spacetime.  Many surprises 
are hidden there.\cite{HNSS, SH}


The spectrum of various fields in the RS spacetime has been analyzed by 
many authors.\cite{GP, Chang}  
Each field has a Kaluza-Klein tower, which has a spectrum
$m_n \sim M_{KK} n$ for large $n$ with the Kaluza-Klein mass scale
$M_{KK}$given by
\beeq
M_{KK} \sim \frac{\pi k}{e^{\pi kR} - 1} 
=\begin{cases}
1/R &\text{as $k \go 0$,}\cr
\pi k e^{-\pi kR}  &\text{for $kR > 2$.}
\end{cases}
\label{KKscale}
\eneq
It is legitimate to suppose that $k= O(M_{\rm Pl})$.  For $kR \sim 12$,
$M_{KK}$ comes out in the TeV range.    In the RS spacetime, the 
spectrum is not with an equal spacing for small $n$.

\section{$W$ boson and $Z$ boson}

As in the flat spacetime,  $W$ bosons and $Z$ bosons acquire finite
masses as the Yang-Mills AB phase $\theta_H$ becomes nonvanishing.
In the $SU(3)$ model the eigenstate of the  $W$ boson becomes a mixture
of $A^{21}_\mu$ and $A^{31}_\mu$.  Its mass is given by
\beeq
m_W \sim \sqrt{\frac{2k}{\pi R}} ~ e^{-\pi kR}
 \Big|  \sin \frac{\theta_H}{2} \Big|~.
  \label{Wmass2}
  \eneq
The neutral current sector is not realistic at all.

In the $SO(5) \times U(1)_{B-L}$ model we denote $SO(5)$ gauge fields
in the $SU(2)_L$, $SU(2)_R$, and $SO(5)/SU(2)_L \times SU(2)_R$
parts by $A^{j_L}_\mu$, $A^{j_R}_\mu$, and $A^{\hat a}_\mu$
($j=1,2,3$, $a= 1,2,3, 4$), and $U(1)_{B-L}$ gauge fields by
$B_\mu$, respectively. 
The  $W$ boson becomes a mixture of $A^{1_L}_\mu + i A^{2_L}_\mu$,
$A^{1_R}_\mu + i A^{2_R}_\mu$, and 
$A^{\hat 1}_\mu + i A^{\hat 2}_\mu$ with a mass
\beeq
m_W \sim \sqrt{ \frac{k}{\pi R} } ~ e^{-\pi kR}
   \big| \sin\theta_H \big|  ~.
  \label{Wmass3}
  \eneq
The  $Z$ boson becomes a mixture of $A^{3_L}_\mu$, $B_\mu$,
$A^{3_R}_\mu$ and $A^{\hat 3}_\mu$ with a mass 
\beqn
&&\hskip -1cm
m_Z \sim \frac{m_W}{\cos \theta_W} ~~, \cr
\noalign{\kern 10pt}
&&\hskip -1cm
\sin \theta_W = \frac{g_B}{\sqrt{ g_A^2 +2  g_B^2}}
= \frac{g_Y}{\sqrt{ g_A^2 + g_Y^2}} 
\label{Zmass1}
\eeqn
where $g_A$ and $g_B$ are the $SO(5)$ and $U(1)$ gauge coupling 
constants, respectively.  $g_Y$ is the weak hypercharge gauge coupling 
constant. 

When $\sin\theta_H$ is $O(1)$, or unless $\theta_H \ll 1$, the relation
(\ref{Wmass3}) implies that $kR \sim 12 ~(6)$ for $k \sim O(M_{\rm Pl})$
($10^{12}$ GeV).   With $kR=12$, $M_{KK}$ turns out $1.8 \sim 3.5$ TeV
for $\theta_H = (0.4 \sim 0.2) \pi$.

\section{Higgs boson}

The 4D Higgs field corresponds to four-dimensional fluctuations of the
Yang-Mills AB phase $\theta_H$ so that its mass and self-couplings 
can be obtained from the effective potential for $\theta_H$.  
In the $SO(5) \times U(1)_{B-L}$ model the zero mode of 
$A_y$ is related to the 4D neutral Higgs field $\phi^0$ by
\beeq
A_y = \sqrt{ \frac{k}{2(e^{2\pi kR}-1)}} ~ e^{2ky} 
~ \phi^0 (x)   \cdot \Lambda ~~.
\label{Higgs2}
\eneq

The effective potential $V_{\eff}(\theta_H)$ at one loop 
is given by
\beeq
V_{\eff}(\theta_H) = \sum \mp \frac{i}{2} 
\int \frac{d^4 p}{(2\pi)^4} \sum_n
\ln \Big\{  p^2 + m_n^2 (\theta_H) \Big\}  ~~.
\label{effV1}
\eneq
It is shown that the $\theta_H$-dependent part of $V_{\eff}(\theta_H)$
is finite, independent of the cutoff scale.\cite{YH1, YH2, Oda1}  
In a model with standard matter content it is
\beeq
V_{\eff}(\theta_H) \sim \frac{3}{128 \pi^6} ~ 
M_{KK}^4 ~ f(\theta_H) ~~,
\label{effV2}
\eneq
where the amplitude of $f(\theta_H)=f(\theta_H + 2\pi)$ is $O(1)$
as confirmed in various models.


Suppose that the global minimum of  $V_{\eff}(\theta_H)$ is located
at $\theta_H \not= 0 ~ (mod~ \pi)$ so that the electroweak 
symmetry breaks down.  By expanding the effective potential around the 
global minimum one can determine the mass of the Higgs boson and
its self-couplings.  The mass and quartic coupling are  found, 
in the $SO(5) \times U(1)_{B-L}$ model,   to be
\beqn
&&\hskip -1cm 
m_H \sim \sqrt{ \frac{3 \alpha_W}{32\pi} f^{(2)}(\theta_H) }
~ \frac{\pi kR}{2} ~ \frac{\sqrt{2} \,  m_W}{\sin \theta_H} ~~, \cr
\noalign{\kern 10pt}
&&\hskip -1cm
\lambda \sim \frac{\alpha_W^2}{16} ~ f^{(4)}(\theta_H) 
\bigg( \frac{\pi kR}{2} \bigg)^2 ~~.
\label{HiggsMass1}
\eeqn
Notice the presence of the enhancement factor $\pi kR/2 \sim 20$, 
which is absent when evaluated in the flat spacetime.  
For $\theta_H = (0.2 \sim 0.5) \pi$, one finds that $m_H = (120 - 210)$ GeV
and $\lambda \sim 0.3$.    Note that in flat spacetime  the values are
$m_H \sim 10$ GeV and $\lambda \sim 0.0008$, which already contradicts
with observation.   

It is surprising that the Higgs mass turns out to be at LHC energies,
though there exists ambiguity in $f^{(2)} (\theta_H)$.  
The enhancement factor originating from the curved space is essential.

\section{Quarks and leptons}

The Lagrangian density for quarks and leptons in a generic form is given by
\beeq
{\cal L}_{f} = 
\psibar i \Gamma^a {e_a}^M \Big\{ \dd_M 
+ \frac{1}{8} \omega_{bcM} [ \Gamma^b ,  \Gamma^c ] 
- ig_A A_M -  \frac{i}{2} g_B q_{B-L}  B_M  \Big\} \psi
 - c  k \, \ep(y) \psibar \psi 
 \label{fermion1}
 \eneq
where  $\ep(y+2\pi R) = \ep(y)= -\ep(-y)$ and $\ep(y)=  1$ for $0<y< \pi R$.
The last term is called a bulk kink mass.\cite{GP}   The dimensionless parameter $c$ 
plays an important role in determining  wave functions of fermions.

Let us consider a fermion multiplet in the spinor representation of $SO(5)$
in the $SO(5) \times U(1)_{B-L}$ model.  
The boundary condition matrices in (\ref{BC2}) are given by
$P_0 = P_1 = \diag (1, 1, -1, -1)$.  $\psi$ contains
\beeq
\psi = \mypmatrix{q_L & q_R \cr Q_L & Q_R}
~~:~~ 
 \mybmatrix{(+,+) & (-,-) \cr (-,-) & (+,+)}
\label{fermion2}
\eneq
where $q$ and $Q$ belong to ({\bf 2}, {\bf 1}) and  ({\bf 1}, {\bf 2})
of $SU(2)_L \times SU(2)_R$, respectively.
$q_L$ and $Q_R$ are even under parity, and have zero modes in the
absence of $A_M$ irrespective of the value of $c$.  

When $\theta_H \not= 0$,  the gauge coupling 
$ g_A \psibar  \Gamma^5 {e_5}^y A_y \psi$  mixes $q$ and $Q$.
Further $A_y(y)$ has nontrivial $y$-dependence in the RS spacetime
so that the mixing with KK excited states also results.  The fermion
mass in four dimensions is determined by finding eigenstates under
such mixing.  

\begin{figure}
\centering  \leavevmode
\includegraphics[width=6.cm]{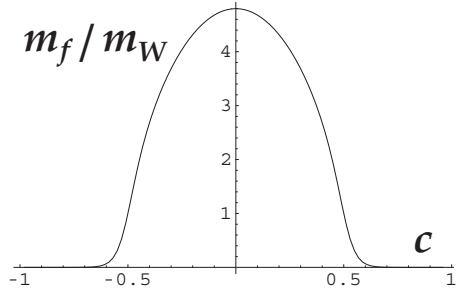}
\caption{The lightest mass eigenvalue~$m_f$ as a function of $c$
at $\theta_W=\onehalf \pi$. 
The vertical axis is in the unit of $m_W$, or 
the fermion mass value at $c=\pm \frac{1}{2}$.}
\label{fmass1}
\end{figure}

To good accuracy the lightest mass eigenvalue is given by
\beeq
 m_f \sim k  \bigg( 
 \frac{c^2 - \frac{1}{4}}
        {e^{\pi kR} \sinh \big[ (c+\half)  k\pi R \big] 
         \,   \sinh \big[ (c-\half) k\pi R \big] }
  \bigg)^{1/2} ~ | \sin \onehalf \theta_H | ~.
  \label{fermionMass}
\eneq
The result is depicted in fig.\ \ref{fmass1} for $\theta_H = \onehalf \pi$.  
$c=1$ corresponds to $m_f = m_W$.  Given the fermion mass $m_f$,
the value of $c$ is determined.   

The remarkable fact is that the values of  $c$ are distributed in the
range $0.43 < c < 0.87$.  The huge hierarchy in the quark-lepton masses 
is explained by  standard distribution of $c$, which is a natural 
quantity in the RS spacetime.  The mass becomes exponentially small
for $c>0.6$.

\begin{table}
\begin{center}
\begin{tabular}{ccccccc}
\hline
& $m_e$ & $m_\mu$ & $m_\tau$ & $m_u$ & $m_c$ & $m_t$ \\
\hline
$c$ & 0.87 & 0.71 & 0.63    & 0.81 & 0.64 & 0.43\\
\hline
\end{tabular}
\end{center}
\caption{The values of $c$ for leptons and quarks}
\label{table1}
\end{table}

\subsection{Suppressed Yukawa couplings}

By inserting the wave functions of the 4D Higgs field and fermions into
$ g_A \psibar  \Gamma^5 {e_5}^y A_y \psi$ and integrating over
$y$, one finds the Yukawa coupling in four dimensions.  In the standard 
model the Yukawa coupling is proportional to the mass of the fermion.  
The relation in the dynamical gauge-Higgs unification is modified, 
becoming  
\beeq
 y_\psi \sim
  \frac{g_A \sqrt{k(c^2 - \frac{1}{4})}}{2e^{\pi kR(c - \half)}} \cdot 
 \cos  \frac{\theta_H}{2}
 =\frac{g m_f}{2m_W}  \cdot  \cos^2 \frac{\theta_H}{2} ~~.
\label{Yukawa1}
\eneq
It is suppressed by a factor $ \cos^2 \onehalf \theta_H $.

\section{Gauge couplings of quarks and leptons}

The electric charge is conserved and the electromagnetic coupling
is universal.  It is the same to all charged particles.  The weak coupling
constants, however, may not be universal once $SU(2)_L \times U(1)_Y$
breaks down to $U(1)_\EM$.   In the standard model those weak couplings
are universal at least at the tree level.  In the dynamical gauge-Higgs
unification small  deviation results.

\begin{figure}[t]
\centering  \leavevmode
\includegraphics[width=7.cm]{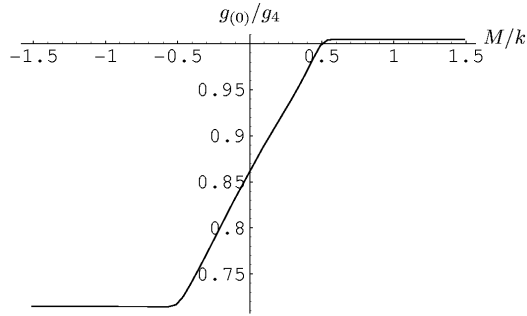}
\caption{The 4D gauge coupling~$g_{(0)}/g_4$ as a function of $c=M/k$
for $\theta_W=\onehalf \pi$ and $kR=12$ in the $SU(3)$ model.}
\label{g0-c}
\end{figure}

Each fermion multiplet couples to the $W$ boson with $g_{(0)} (\theta_H, c)$
obtained by integrating over $y$ with wave functions of  $W$ and fermions inserted
in the gauge interaction term in (\ref{fermion1}).
$g_{(0)}$ depends on both $\theta_H$ and $c$.  It is depicted in 
fig.\ \ref{g0-c} as a function of $c=M/k$ at $\theta_H= \onehalf \pi$.  For $c>\onehalf$
the deviation is very small.    The asymptotic value for $c < \onehalf$ is $\cos\theta_H$
in the $SU(3)$ model.

For the values of $c (> 0.43)$ for quarks and leptons in table 1, the dependence
of $g_{(0)}$ on $c$ is small.  The violation of the $\mu$-$e$, $\tau$-$e$, and
$t$-$e$ universality in the charged current interactions is of order of $10^{-8}$, 
$2 \times 10^{-6}$  and $2 \times 10^{-2}$, respectively.


Each quarks and leptons couples to the KK excited states of gauge bosons 
as well.  It was noticed that those couplings can be large at $c=0$, which
gives rise to contradition with observation unless $M_{KK}$ is sufficiently
large.

The coupling $g_{(n)}$ of a fermion to the $n$-th KK excited state of $W$ is
depicted in fig.\ 3 at $\theta_H = \onehalf$.  It is seen that the couplings are
very small for $c>\onehalf$ as noted by Gherghetta and Pomarol so that
the earlier constgraint on $M_{KK}$ is evaded.

\begin{figure}[t]
\centering  \leavevmode
\includegraphics[width=8.cm]{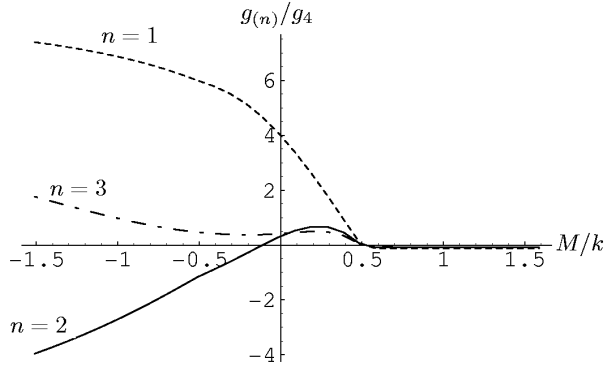}
\caption{The gauge couplings to the $n$-th KK excited states of 
$W$, $g_{(n)}$ ($n=1,2,3$) as functions of $c = M/k$ 
at $\theta_W=\onehalf\pi$ in the $SU(3)$ model. }
\label{gn-c}
\end{figure}

\section{$WWZ$, $WWH$, and $ZZH$ couplings}

At $\theta_H \not= 0$, an eigenstate of each field becomes mixture of various
components of the multiplet which  the field belong to.  All of the $W$, $Z$, and $H$ 
fields in four dimensions are parts of the gauge field multiplets.   The mixing pattern
is not identical among these fields so that the effective 4D couplings necessarily
depend on $\theta_H$.  This would provide critical tests for the dynamical 
gauge-Higgs unification particularly in the $WWZ$, $WWH$, and $ZZH$ couplings.

The $WWZ$ coupling is found to be
\beeq
g_{WWZ} \simeq g \cos \theta_W ~.
\label{WWZ1}
\eneq
To this order the coupling is the same as  in the standard model.  
The experiment at LEP2 indicates the validity of the standard model.

The $WWH$ and $ZZH$ couplings are found to be
\beqn
&&\hskip -1cm
\lambda_{WWH} \simeq g m_W \cdot  p_H \cos \theta_H ~~, \cr
\noalign{\kern 5pt}
&&\hskip -1cm
\lambda_{ZZH} \simeq \frac{g m_Z}{\cos \theta_W} \cdot  p_H \cos \theta_H  ~~,
\label{WWH1}
\eeqn
where $p_H = {\rm sign}(\sin \theta_H )$.  These couplings are suppressed by a
factor $\cos \theta_H$. Note that these couplings are important in drawing a
constraint for the Higgs boson mass from the LEP data as well.



\newpage

\leftline{\bf Acknowledgments}

\vskip .2cm 

The author would like to thank the Aspen Center for Physics for its 
hospitality where a part of this work was performed.
This work was supported in part  by  Scientific Grants from the Ministry of 
Education and Science, Grant No.\ 17540257,
Grant No.\ 13135215 and Grant No.\ 18204024.


\def\jnl#1#2#3#4{{#1}{\bf #2} (#4) #3}

\def\Zphys{{\em Z.\ Phys.} }
\def\jssc{{\em J.\ Solid State Chem.\ }}
\def\jpsJ{{\em J.\ Phys.\ Soc.\ Japan }}
\def\ptps{{\em Prog.\ Theoret.\ Phys.\ Suppl.\ }}
\def\PTP{{\em Prog.\ Theoret.\ Phys.\  }}

\def\JMP{{\em J. Math.\ Phys.} }
\def\NPB{{\em Nucl.\ Phys.} B}
\def\NP{{\em Nucl.\ Phys.} }
\def\PLB{{\em Phys.\ Lett.} B}
\def\PL{{\em Phys.\ Lett.} }
\def\PRL{\em Phys.\ Rev.\ Lett. }
\def\PRB{{\em Phys.\ Rev.} B}
\def\PRD{{\em Phys.\ Rev.} D}
\def\PRe{{\em Phys.\ Rep.} }
\def\AP{{\em Ann.\ Phys.\ (N.Y.)} }
\def\RMP{{\em Rev.\ Mod.\ Phys.} }
\def\ZPC{{\em Z.\ Phys.} C}
\def\SCI{\em Science}
\def\CMP{\em Comm.\ Math.\ Phys. }
\def\MPLA{{\em Mod.\ Phys.\ Lett.} A}
\def\IJMPA{{\em Int.\ J.\ Mod.\ Phys.} A}
\def\IJMPB{{\em Int.\ J.\ Mod.\ Phys.} B}
\def\EPJC{{\em Eur.\ Phys.\ J.} C}
\def\PR{{\em Phys.\ Rev.} }
\def\JHEP{{\em JHEP} }
\def\cmp{{\em Com.\ Math.\ Phys.}}
\def\JPA{{\em J.\  Phys.} A}
\def\JPG{{\em J.\  Phys.} G}
\def\NJP{{\em New.\ J.\  Phys.} }
\def\CQG{\em Class.\ Quant.\ Grav. }
\def\ATMP{{\em Adv.\ Theoret.\ Math.\ Phys.} }
\def\ibid{{\em ibid.} }

\renewenvironment{thebibliography}[1]
         {\begin{list}{[$\,$\arabic{enumi}$\,$]}  
         {\usecounter{enumi}\setlength{\parsep}{0pt}
          \setlength{\itemsep}{0pt}  \renewcommand{\baselinestretch}{1.2}
          \settowidth
         {\labelwidth}{#1 ~ ~}\sloppy}}{\end{list}}

\def\reftitle#1{}                


\vskip .8cm

\leftline{\bf References}

\begin{thebibliography}{9}
\small
\baselineskip=13pt

\bibitem{YH1}
Y.\ Hosotani, \jnl{\PLB}{126}{309}{1983}.
\reftitle{Dynamical Mass Generation By Compact Extra Dimensions}

\bibitem{YH2}
Y.\ Hosotani, \jnl{\AP}{190}{233}{1989}.
\reftitle{Dynamics Of Nonintegrable Phases And Gauge Symmetry Breaking}

\bibitem{Fairlie1}
D.B.\ Fairlie, \jnl{\PLB}{82}{97}{1979};
\reftitle{Higgs' Fields And The Determination of The Weinberg Angle}
\jnl{\JPG}{5}{L55}{1979}.
\reftitle{Two Consistent Calculations of The Weinberg Angle}

\bibitem{Manton1}
N.\ Manton, \jnl{\NPB}{158}{141}{1979}.
\reftitle{A New Six-Dimensional Approach to the Weinberg-Salam Model}

\bibitem{Lim2}
H.\ Hatanaka, T.\ Inami and C.S.\ Lim, 
\jnl{\MPLA}{13}{2601}{1998}.
\reftitle{The Gauge Hierarchy Problem and Higher Dimensional Gauge Theories}

\bibitem{Antoniadis1}
I.\ Antoniadis, K.\ Benakli and M.\ Quiros,
\jnl{\it New. J.\ Phys.}{3}{20}{2001}.
\reftitle{Finite Higgs mass without Supersymmetry}

\bibitem{Lim1}
M.\ Kubo, C.S.\ Lim and H.\ Yamashita,
 \jnl{\MPLA}{17}{2249}{2002}.
\reftitle{The Hosotani Mechanism in Bulk Gauge Theories with an Orbifold Extra   Space $S^1/Z_2$}



\bibitem{Csaki1}
C.\ Csaki, C.\ Grojean and H.\ Murayama, \jnl{\PRD}{67}{085012}{2003};
\reftitle{Standard Model Higgs From Higher Dimensional Gauge Fields}

C.A.\ Scrucca, M.\ Serone and L.\ Silverstrini, \jnl{\NPB}{669}{128}{2003}. 
\reftitle{Electroweak symmetry breaking and fermion masses from extra dimensions}

\bibitem{gaugeHiggs3}
L.J.\ Hall, Y.\ Nomura and D.\ Smith,  \jnl{\NPB}{639}{307}{2002};
\reftitle{Gauge-Higgs Unification in Higher Dimensions}

L.\ Hall, H.\ Murayama, and Y.\ Nomura, 
   \jnl{\NPB}{645}{85}{2002};
\reftitle{Wilson Lines and Symmetry Breaking on Orbifolds}

G.\ Burdman and Y.\ Nomura, \jnl{\NPB}{656}{3}{2003}; 
\reftitle{Unification of Higgs and Gauge Fields in Five Dimensions}



C.A.\ Scrucca, M.\ Serone, L.\ Silvestrini and A.\ Wulzer,
\jnl{\JHEP}{0402}{49}{2004}.
\reftitle{Gauge-Higgs Unification in Orbifold Models} 


\bibitem{HHHK}
N.\ Haba, M.\ Harada, Y.\ Hosotani and Y.\ Kawamura, 
\jnl{\NPB}{657}{169}{2003};   
{\it Erratum}, {\it ibid.}  B{\bf 669} (2003) {381}.
\reftitle{Dynamical Rearrangement of Gauge Symmetry on the Orbifold $S^1/Z_2$}


\bibitem{HHKY}
N.\ Haba,  Y.\ Hosotani,  Y.\ Kawamura and T.\ Yamashita, 
\jnl{\PRD}{70}{015010}{2004};
\reftitle{Dynamical symmetry breaking in Gauge-Higgs unification on orbifold}

N.\ Haba,  K.\ Takenaga, and T.\ Yamashita, 
\jnl{\PLB}{615}{247}{2005}.
\reftitle{Higgs mass in the gauge-Higgs unification}

\bibitem{HNT2}
Y.\ Hosotani, S.\ Noda and K.\ Takenaga,
\jnl{\PLB}{607}{276}{2005}.
\reftitle{Dynamical Gauge-Higgs Unification in the Electroweak Theory}

\bibitem{Csaki2}
G.\ Cacciapaglia, C.\ Csaki and S.C.\ Park,
\jnl{\JHEP}{0603}{099}{2006}.
\reftitle{Fully Radiative Electroweak Symmetry Breaking}

\bibitem{Panico2}
G.\ Panico, M.\ Serone and A.\ Wulzer,
\jnl{\NPB}{739}{186}{2006}.
\reftitle{A Model of Electroweak Symmetry Breaking from a Fifth Dimension}


\bibitem{Pomarol2}
R.\ Contino, Y.\ Nomura and A.\ Pomarol, \jnl{\NPB}{671}{148}{2003}.
\reftitle{Higgs as a holographic pseudo-Goldstone boson}

\bibitem{Agashe2}
K.\ Agashe, R.\ Contino and A.\ Pomarol, 
\jnl{\NPB}{719}{165}{2005}.
\reftitle{The minimal composite Higgs model}


\bibitem{Oda1}
K.\ Oda and A.\ Weiler, \jnl{\PLB}{606}{408}{2005}.
\reftitle{Wilson Lines in Warped Space: Dynamical Symmetry Breaking and Restoration}


\bibitem{HM}
Y.\ Hosotani and M.\ Mabe, \jnl{\PLB}{615}{257}{2005}.
\reftitle{Higgs boson mass and electroweak-gravity hierarchy
from dynamical gauge-Higgs unification in the warped spacetime}


\bibitem{HNSS}
Y.\ Hosotani, S.\ Noda, Y.\ Sakamura and S.\ Shimasaki, 
\jnl{\PRD}{73}{096006}{2006}.
\reftitle{Gauge-Higgs Unification and Quark-Lepton Phenomenology 
in the Warped Spacetime}

\bibitem{Carena}
M.\ Carena, E.\ Ponton, J.\ Santiago and C.E.M.\ Wagner,
hep-ph/0607106.
\reftitle{Light Kaluza Klein States in Randall-Sundrum Models with Custodial SU(2)}

\bibitem{SH}
Y.\ Sakamura and Y.\ Hosotani, 
hep-ph/0607236.
\reftitle{WWZ, WWH, and ZZH Couplings in the Dynamical 
Gauge-Higgs Unification in the Warped Spacetime}


\bibitem{YHscgt2}
Y.\ Hosotani, in the Proceedings of 
{\it ``Dynamical Symmetry Breaking"},  ed. M. Harada and K. Yamawaki 
(Nagoya University, 2004), p.\ 17. (hep-ph/0504272).
\reftitle{Dynamical Gauge Symmetry Breaking by Wilson Lines
in the Electroweak Theory}

\bibitem{Irges}
N.\ Irges and F.\ Knechtli, 
\jnl{\NPB}{719}{121}{2005};  hep-lat/0604006.

\bibitem{Maru1}
N.\ Maru and T.\ Yamashita, hep-ph/0603237.
\reftitle{Two-loop calculation of Higgs Mass in gauge-Higgs unification:
5d Massless QED compactified on $S^1$}

\bibitem{YHfinite}
Y.\ Hosotani, hep-ph/0607064.
\reftitle{All-order finiteness of the Higgs boson mass in the 
dynamical gauge-Higgs unification}

\bibitem{RS1}
L.\ Randall and R.\ Sundrum,  \jnl{\PRL}{83}{3370}{1999}.


\bibitem{GP}
T.\ Gherghetta and A.\ Pomarol,
\jnl{\NPB}{586}{141}{2000}.
\reftitle{Bulk fields and supersymmetry in a slice of AdS}

\bibitem{Chang}
S.\ Chang, J.\ Hisano, H.\ Nakano, N.\ Okada and M.\ Yamaguchi, 
\jnl{\PRD}{62}{084025}{2000}.
\reftitle{Bulk Standard Model in the Randall-Sundrum Background}





\end{thebibliography}
\end{document}